\documentclass[aps,floats,superscriptaddress,twocolumn,pre,showpacs]{revtex4-1}

\usepackage{graphicx,epsfig}% Include figure files
\usepackage{graphics,dcolumn,bm,epic, eepic,fleqn,float}
\usepackage{amssymb,amsmath,amsfonts,multirow,rotate,color}
\usepackage{soul}
\usepackage{wasysym}
\usepackage{dcolumn}
\usepackage{bm}

\definecolor{Myorange}{cmyk}{0,0.42,1,0}

\newcommand{\lay}[1]{^{[#1]}}

\begin{document}

\title{The new challenges of multiplex networks: measures and models}

\author {Federico Battiston}
\affiliation{School of Mathematical Sciences, Queen Mary University of London, 
London E1 4NS, United Kingdom}   

\author {Vincenzo Nicosia}
\affiliation{School of Mathematical Sciences, Queen Mary University of London, 
London E1 4NS, United Kingdom}

\author{Vito Latora}
\affiliation{School of Mathematical Sciences, Queen Mary University of London, 
London E1 4NS, United Kingdom}  

\affiliation{Dipartimento di Fisica ed Astronomia, Universit\`a di Catania and INFN, I-95123 Catania, Italy}

\date{\today}
%ABSTRACT
\begin{center}
\begin{abstract}
What do societies, the Internet, and the human brain have in common? They are all examples of complex relational systems, whose emerging behaviours are largely determined by the non-trivial networks of interactions among their constituents, namely individuals, computers, or neurons, rather than the properties of the units themselves. In the last two decades, network scientists have proposed models of increasing complexity to better understand real-world systems. Only recently we have realised that multiplexity, i.e. the coexistence of several types of interactions among the constituents of a complex system, is responsible for substantial qualitative and quantitative differences in the type and variety of behaviours that a complex system can exhibit. As a consequence, multilayer and multiplex networks have become a hot topic in complexity science. Here we provide an overview of some of the measures proposed so far to characterise the structure of multiplex networks, and a selection of models aiming at reproducing those structural properties and quantifying their statistical significance. Focusing on a subset of relevant topics, this brief review is a quite comprehensive introduction to the most basic tools for the analysis of multiplex networks observed in the real-world. The wide applicability of multiplex networks as a framework to model complex systems in different fields, from biology to social sciences, and the colloquial tone of the paper will make it an interesting read for researchers working on both theoretical and experimental analysis of networked systems.
\end{abstract}
\end{center}

\maketitle

\section{Introduction}
\label{intro}
One of the most intriguing characteristic of complex systems is that
most of the collective behaviours they exhibit cannot be predicted
from the the knowledge of the properties of their elementary
constituents. Indeed, in the last two decades network science has
shown that in many cases, from biology to economics, the structure of
the interactions among the constituents of the system plays a
fundamental role in shaping the emergence of complex behaviours, much
more important than the role played by the specific properties of the
single units of the
system~\cite{barabasi02,newman03,Boccaletti2006,newman10}. And
surprisingly, systems as diverse as social networks, transportation
systems, cities, and the human brain were shown to share a significant
number of features and a comparable structure of
interactions\cite{strogatz01,arenas08,bullmore09,castellano09,barthelemy11,fallani14}.

Recently, the availability of new data sets, the rediscovery of old
ones and the access to more powerful computers, has highlighted the
necessity to develop a new framework to represent networks whose units
interact through more than just one type of relations. These systems
are usually called multiplex networks, and are characterised by the
fact that all the connections of a given type are embedded into a
distinct layer. A full description of early research on such topic can
be found in~\cite{boccaletti14,kivela14,lee15}. This Article provides
an informal, still comprehensive handbook for the experimental
investigation of systems which can be described as multiplex
networks. In the first part we provide an overview of the most basic
measures to characterise the structure of multiplex networks, focusing
on the properties of nodes, edges, and layers. In the second part we
review a few models which can be used to reproduce the empirical
patterns observed in real-world multi-layer system, or to assess the
statistical significance of such patterns.

%%%%%%%%%%%%%%%%                     MEASURES            %%%%%%%%%%%%%%%%%%%%%

\section{Measures for multiplex networks}
\label{sec:structure}
We consider a multiplex network $\mathcal M$ consisting of $N$ nodes
and $M$ different types of relations, represented by $M$ graphs, or
layers. We can fully describe the structure of the system by
considering the set of adjacency matrices
\begin{equation}
  \mathcal M \equiv \bm{A}=\{A\lay{1}, \ldots, A\lay{M}\},
  \label{eq:multiplex}
\end{equation}
where $A\lay{\alpha}=\{ a_{ij}\lay{\alpha}\}$, with
$a_{ij}\lay{\alpha}=1$ if $i$ and $j$ share a bond of type $\alpha$
and $a_{ij}\lay{\alpha}=0$ otherwise~\cite{battiston14}. When the
connections among nodes are weighted, the system can be described by a
set of weighted adjacency matrices $\bm{W}=\{W\lay{1}, \ldots,
W\lay{M}\}$, with $W\lay{\alpha}=\{ w_{ij}\lay{\alpha}\}$ and
$w_{ij}\lay{\alpha}$ the weight of the link between node $i$ and
$j$~\cite{menichetti14}.
This formulation implicitly assumes that each node $i$ consists of $M$
replicas, one at each layer, and that a link can only connect two
replicas lying on the same layer. Consequently, although both $\bm{A}
= \{a_{ij}\lay{\alpha}\}$ and $\bm{W}=\{w_{ij}\lay{\alpha}\}$ can be
considered as generic order-3 tensors, it is important to stress that
node $i$ on layer $\alpha$ and node $i$ on layer $\beta$ effectively
represent the same unit of the system and not two different ones. In
other words, the replicas of the same node are identified across
layers. Social systems can be naturally cast within this framework,
where different layers can represent for instance different
interaction channels among the same nodes (e.g., face-to-face
communication, email exchange, online chat, etc.), but the different
replicas are just a mathematical representation of the same individual
in each of the $M$ contexts.

However, there are cases in which there exists some sort of
communication or flow between the replicas of the same node at
different layers. A typical example is that of multimodal
transportation systems, where nodes are locations and layers represent
different transport modalities, e.g. bus, underground, trains, etc. In
this case, an accurate modelling of the system has to take into
account inter-layer transitions between the replicas of the same node
at different layers, and it is more convenient to model the structure
of the system through an order-4 tensor
$\mathcal{M}_{i,\alpha}^{j,\beta}$~\cite{dedomenico13}. This
formulation makes explicit (and adjustable) the relative importance of
intra-layer and inter-layer
connections~\cite{superdiffusion,radicchi14}. Unless specified
otherwise, in the following we will mostly use the more simple
formulation based on order-3 tensors and given in
Eq.~\ref{eq:multiplex}.

%%%%%%%% Node properties
\subsection{Node properties}

Differently from the traditional single-layer approach, where node
properties are described by scalar variables, node features in
multiplex networks are naturally described in vectorial terms. As an
example, for each node $i$ we consider its total number of
connections, or degree, at layer $\alpha$,
i.e. $k\lay{\alpha}_i=\sum_{j \neq i}^N a_{ij}\lay{\alpha}$, and the
multilayer degree $\bm{k}_i=\{k_i\lay{1}, \ldots, k_i\lay{M}\}$. A
crucial empirical evidence is that in many multiplex networks not all
nodes have connections at all layers. As a consequence, a node $i$ is
defined as active on a layer $\alpha$ if it is connected to at least
another node at that layer, i.e. if $k_i\lay{\alpha}>0$. The
activity-pattern of each node can be compactly stored into the
node-activity vector
\begin{equation}
\bm{b}_i=\{b_i\lay{1},  \ldots, b_i\lay{M}\},
\end{equation}
where $b_i\lay{\alpha}=1-\delta_{0,k_i\lay{\alpha}}$,
i.e. $b_i\lay{\alpha}=1$ if node $i$ is active on layer $\alpha$, and
$b_i\lay{\alpha}=0$ otherwise. The total activity
$B_i=\sum_{\alpha=1}^M b_i\lay{\alpha}$ represents the number of
layers in which node $i$ is active, with $0\le B_i \le
M$~\cite{nicosia15}. It has been found that most real-world multiplex
networks are characterised by heterogeneous distributions of node
activity~\cite{nicosia15}, and it has been shown that such
heterogeneity might be responsible for the increased fragility of
multiplex networks to random failures~\cite{cellai16}.

Given a generic vectorial property $\bm{\xi}_i=\{\xi_i\lay{1}, \ldots,
\xi_i\lay{M}\}$, it is important to be able to compress the information
into meaningful scalar descriptors, especially for systems composed of
a large number of layers. A typical way to approach this problem is to
consider the first and the second moment of the vector $\bm{\xi}$,
accounting for its mean value $\mu(\bm{\xi})$ (or, analogously, the sum
of its components) and its variance $\sigma^2(\bm{\xi})$, or related
quantities. In the particular case of the degree, the total number of
connections of node $i$ is usually called total or overlapping
degree~\cite{battiston14}
\begin{equation}
o_i=\sum_{i=1}^N k_i \lay {\alpha}
\end{equation}
while the heterogeneity of the number of neighbours of node $i$ across
the layers can be measured through the multiplex participation
coefficient~\cite{battiston14}
\begin{equation}
P_i=\frac{M}{M-1}\left[1-
  \sum_{\alpha=1}^M\biggl(\frac{k_i^{[\alpha]}}{o_i}\biggr)^2\right],
\label{participationcoefficient}
\end{equation}
where $P_i=1$ when the links incident on node $i$ are equally
distributed across the layers, and $P_i=0$ when a node is only active
on one layer. We note that similar information about the heterogeneity
of the distribution of a node's connections across layers is provided by
the Shannon entropy of the degree vector~\cite{battiston14}
\begin{equation}
H_i
=-\sum_{\alpha=1}^M\frac{k_i^{[\alpha]}}{o_i}\ln \biggl(\frac{k_i^{[\alpha]}}{o_i}\biggr).
\end{equation}

The pair of variables $(P_i, o_i)$ can be used to classify nodes via
the so-called multiplex cartography~\cite{battiston14}, efficiently
distinguishing multiplex hubs (high $o_i$ and high $P_i$), focused
hubs (high $o_i$ and low $P_i$), multiplex leaves (low $o_i$ and high
$P_i$) and focused leaves (low $o_i$ and low $P_i$).

A remarkable property of real-world networks is the tendency of nodes
to form triangles, a phenomenon usually known as transitivity. In
single-layer networks, the abundance of triangles is typically
measured through the average clustering coefficient $C$, where
$C=\frac{1}{N}\sum_{i=1}^N C_i$ and $C_i$ accounts for the fraction of
triads centred on node $i$ which are closed into triangles. In
multiplex networks, triads and triangles can effectively extend over
more than one layer. We define an m-triad (m-triangle) a triad
(triangle) which uses edges from $m$ different layers. It is possible
to define two multiplex clustering coefficients to quantify the added
value provided to transitivity by the
layered structure~\cite{battiston14}. For each node $i$, the first
coefficient $C_{i,1}$ is defined as the ratio between the number of
2-triangles with a vertex in $i$ and the number of 1-triads centred
in $i$. In formulas:
\begin{equation}
\label{clustering1eq}
C_{i,1}
= \frac{\sum_{\alpha} \sum_{\alpha' \neq \alpha} \sum_{j \neq i, m \neq i } (a_{ij}^{[\alpha]}a_{jm}^{[\alpha']}a_{mi}^{[\alpha]})}{(M-1)\sum_{\alpha} k_i^{[\alpha]}(k_i^{[\alpha]}-1)}
\end{equation}
The second multiplex clustering coefficient $C_{i,2}$ is instead
defined as the ratio between the number of 3-triangles with node $i$
as a vertex, and the number of 2-triads centred in $i$. In formulas:
\begin{equation}
\label{clustering2eq}
C_{i,2}= \frac{\sum_{\alpha} \sum_{\alpha' \neq \alpha} \sum_{\alpha''
    \neq \alpha, \alpha'}\sum_{j \neq i, m \neq i}
  (a_{ij}^{[\alpha]}a_{jm}^{[\alpha'']}a_{mi}^{[\alpha']}) }
{(M-2)\sum_{\alpha} \sum_{\alpha' \neq \alpha}  \sum_{j \neq i, m \neq i }  
(a_{ij}^{[\alpha]}a_{mi}^{[\alpha']})}.
\end{equation}
These two measures are defined respectively for $M \ge 2$ and $M \ge
3$, and are a natural generalisation of clustering coefficient
to the case of multiplex networks. A related generalisation of
clustering coefficient, based on the order-4 tensorial formulation for
multiplex networks, has been suggested in Ref.~\cite{cozzo15}.

Another characteristic properties of real-world networks is the
presence of heterogeneity in the relative importance of nodes, as
measured by different notions of node
centrality~\cite{Boccaletti2006}.  A number of different approaches
have been suggested to define and compute the centrality of a node in
a multiplex network. A first possibility consists in defining the
multiplex centrality as a combination of the centrality scores of each
node at the different layers. For instance, starting from the
centrality vector of node $i$, $\bm {c}_i=\{c_i\lay{1}, \ldots,
c_i\lay{M} \}$, one can try to condense the information into a single
scalar variable, as is normally done for the degree. However,
computing averages of centrality scores across layers is not always
meaningful. The first reason is that in general a node can play
different roles on different layers, and averaging over layers will
only level down such heterogeneities.

The presence of more than one layer allows to define new genuinely
multiplex centrality measures in which the role of a node explicitly
depends on the structure of the multiplex at all layers. For instance,
the authors of Ref.~\cite{sola13} suggested to compute the eigenvector
centrality of nodes on each layer $\alpha$ as the normalised
eigenvector relative to the largest eigenvalue of
\begin{equation}
\tilde A\lay{\alpha}=\sum_{\beta=1}^M i^{[\alpha,\beta]} A\lay{\beta}
\end{equation}
where $I=\{ i^{[\alpha,\beta]} \}$ is a given influence matrix which
determines how the centrality of layer $\alpha$ depends on the
structure of layer $\beta$. In Ref.~\cite{battiston14}, instead, the
authors studied the contributions of the different layers to the
centrality of the nodes by varying the coefficients $i\lay{\alpha}$
$\alpha=1,\ldots,M$ of the matrix
\begin{equation}
A'=\sum_{\alpha=1}^M i^{[\alpha]} A\lay{\alpha}
\end{equation}
which is a convex combination of the adjacency matrices of the layers.

An entire class of node centrality measures can be defined by using
the properties of random walks on multiplex
networks~\cite{battiston16rw,sole16rw}. A particularly interesting
example is that of multiplex PageRank centrality proposed in
Ref.~\cite{halu13}. The authors of Ref.~\cite{halu13} considered the
case of a two-layer multiplex network and defined the multiplex
PageRank of the nodes in layer $\alpha=2$ as a function of the
PageRank scores of the nodes in layer $\alpha=1$. The main idea of
this genuinely multiplex measure is that, especially in social
systems, nodes can leverage their centrality in one context, such as
personal relationships (represented by layer $1$) to gain centrality
in another context, e.g. professional relationships (represented by
layer $2$).

Real-world networks often exhibit the small-world property, meaning
that the typical distance between any pair of nodes in the system
scales logarithmically with the total number of nodes. An important
observation is that not all the nodes of a system are equally
important in mediating paths between other nodes, which is the main
idea between the concept of node betweenness~\cite{Boccaletti2006}.
In a multiplex system, the reachability of a node might significantly
depend on the interplay between different layers. The added value
introduced by multiplexity can be measured through the
interdependence~\cite{morris12,nicosia13}
\begin{equation}
\lambda_i = \frac{1}{N-1}\sum_{j \neq i}
\frac{\psi_{ij}}{\sigma_{ij}},
\end{equation}
where $\psi_{ij}$ accounts for the number of shortest paths between
$i$ and $j$ that use edges in more than one layer, and $\sigma_{ij}$
is the total number of shortest paths between $i$ and $j$. The
quantity $\lambda_i$ takes values in the interval $[0,1]$, with larger
values corresponding to a higher advantage for the reachability of
node $i$ provided by the interplay of the different layers. By averaging over all nodes we obtain the interdependence of the multi-layer system $\lambda=1/N \sum_{i}\lambda_i$. It is also possible to define a layer interdependence $\lambda \lay \alpha$, accounting for the number of shortest paths with at least one link on layer $\alpha$~\cite{aleta16}.

Finally, if one represents a multiplex network using an order-4
tensor, an entire class of centrality measures can be obtained as
natural extensions to adjacency tensors of the corresponding measures
defined on adjacency matrices~\cite{sole14}. For instance, the
eigenvector centrality of a node in this formalism can be computed by
considering either the eigenvectors of the order-4 tensorial
representation of the multiplex or the eigenvectors of the associated
supra-adjacency matrix. An interesting application of this class of
measures, described in Ref.~\cite{dedomenico15vers,sole14}, allows to define
the versatility of nodes, assigning higher centrality scores to those
nodes which act as bridges among different layers.

%%%%%%%% Layer properties
\subsection{Layer properties}

Similarly to the case of node activity, it is possible to define the
activity-vector of each layer $\alpha$~\cite{nicosia15} as
\begin{equation}
\bm{d}\lay{\alpha}=\{b_1\lay{\alpha},  \ldots, b_N\lay{\alpha}\},
\end{equation}
where $b_i\lay{\alpha}=1$ if $k_i\lay{\alpha}>0$, and
$b_i\lay{\alpha}=0$ otherwise.  For each layer $\alpha$, the total
layer activity $N\lay{\alpha}= \sum_{i=1}^N b_i\lay{\alpha}$ describes
the total number of nodes with at least one connection in layer
$\alpha$, with $0 \le N\lay{\alpha} \le N$. The similarity between the
activity-vectors of two layers $\alpha$ and $\beta$ can be
measured by mean of the pairwise multiplexity~\cite{nicosia15}, which
accounts for the fraction of nodes of the multiplex which are active
on both layers:
\begin{equation}
Q\lay{\alpha, \beta}=\frac{1}{N}\sum_{i=1}^N b_i\lay{\alpha}.
b_i\lay{\beta}.
\end{equation}
In general $0\le Q\lay{\alpha, \beta} \le 1$, with $Q\lay{\alpha, \beta}=1$
when all nodes are active in both layers, and $Q\lay{\alpha, \beta}=0$
when no node is active on both layers.
The similarity among the patterns of activity in two layers can also
be measured through the Hamming distance~\cite{nicosia15}
\begin{equation}
H\lay{\alpha, \beta}=\frac{\sum_i b_i\lay \alpha (1 - b_i\lay \beta) +
  b_i\lay \beta (1 - b_i\lay \alpha)}{\rm
  {min}(N\lay{\alpha}+N\lay{\beta}, N)}.
\end{equation}
where $H\lay{\alpha, \beta}=0$ if $\bm d\lay \alpha = \bm d\lay \beta$
and $H\lay{\alpha, \beta}=1$ if all active nodes are active in no more
than one layer. It has been suggested that real multiplex networks are
normally characterised by heterogeneous distributions of layer
activity and of pairwise multiplexity~\cite{nicosia15}.

Another interesting property observed in real multiplex networks is
the presence of correlations between the degrees of the same node at
different layers. This is normally signalled by the fact that the
probability $P(k\lay{\alpha}=k_1, k\lay{\beta}=k_2)$ to find a node
with degree $k_1$ on layer $\alpha$ and degree $k_2$ on layer $\beta$
does not factorise in the product $P\lay{\alpha}(k)P\lay{\beta}(k)$ of
the degree distributions of the two layers.  In general, given two
layers $\alpha$ and $\beta$ and a generic node property
$\bm{\xi}_i$, the correlation between $\xi\lay{\alpha}_i$ and
$\xi\lay{\beta}_i$ can be computed using the rank correlation
coefficient~\cite{nicosia15}:
\begin{equation}
  \rho \lay{\alpha, \beta} = \frac{\sum_{i}\left(   R\lay{\alpha}_i -
        \overline{ R\lay{\alpha} }    \right) 
\left( R\lay{\beta}_i -
       \overline{ R\lay{\beta} }      \right)}
    {\sqrt{\sum_i\left(   R\lay{\alpha}_i -
        \overline{ R\lay{\alpha} }    \right)^2\sum_j \left(  
 R\lay{\beta}_j -  \overline{ R\lay{\beta} } 
\right)^2}}
\end{equation}
where $R\lay{\alpha}_i$ is the rank of node $i$ at layer $\alpha$
induced by the property $\bm \xi$. When the property of interest is the
degree, it makes sense to define the quantity:
\begin{equation}
  \overline{k\lay{\beta}}(k\lay{\alpha}) =
  \sum_{k\lay{\beta}}k\lay{\beta} P(k\lay{\beta} | k\lay{\alpha})
\end{equation}
that is the average degree at layer $\beta$ of a node having degree
$k\lay{\alpha}$ at layer $\alpha$, and is the multiplex homologous of
the nearest-neighbours average degree function $k_{nn}(k)$
traditionally used to quantify degree-degree correlations in
single-layer graphs~\cite{Pastor2001}. An increasing (decreasing)
trend in $\overline{k\lay{\beta}}(k\lay{\alpha})$ will signal the
presence of positive (negative) inter-layer degree correlations
between layer $\alpha$ and layer $\beta$.

The authors of Ref.~\cite{lacasa15} proposed to quantify inter-layer
degree correlations by using the pairwise mutual information between
the degree sequences of the two layers:
\begin{equation}
I \lay {\alpha,\beta}=\sum_{k\lay{\alpha}}\sum_{k\lay{\beta}}P(k\lay{\alpha},
  k\lay{\beta}) \log{\frac{P(k\lay{\alpha},
      k\lay{\beta})}{P(k\lay{\alpha})P(k\lay{\beta})}}
  \label{eq:MI}
\end{equation}
which is maximal when the degree sequences
$\left\{k_i\lay{\alpha}\right\}$ and $\left\{k_i\lay{\beta}\right\}$
are perfectly correlated (or perfectly anti-correlated), and minimal
when they are uncorrelated.

An equivalent set of quantities to measure the inter-layer
assortativity has been defined for the order-4 tensorial formulation
in Ref.~\cite{ferraz15}.

An analysis of the spectral properties of multiplex networks can be found in~\cite{sole2013,sanchez2014}.

A fundamental research question in the field of multiplex networks is
to assess whether the presence of more than one interaction layer
indeed provides more information about the structure of a system
compared to a classical single-layer network representation. In
particular, it is interesting to quantify how much information is lost
(if any at all) when we aggregate some or all the layers of a
multiplex network to obtain a lower-dimensional representation. The
authors of Ref.~\cite{dedomenico15red} tackled the problem of
multiplex reducibility by drawing on an existing formal parallel
between density operators of quantum systems and Laplacian matrices of
graphs, and extending the concept of Von Neumann entropy of a graph to
the case of multiplex networks. They proposed a greedy procedure,
based on the estimation of the quantum Jensen-Shannon divergence
between layers, which allows to successively aggregate the most
redundant layers of a multiplex and to obtain a more compact
representation which uses the minimal number of layers while
maximising the distinguishability between the multiplex and the
single-layer representation of the same system. An interesting result
of the paper is that different multilayer systems allow different
levels of reducibility, with man-made systems being the least
reducible and biological and social systems showing the highest levels
of redundancy~\cite{dedomenico15red}. 

Interestingly, many real-world multiplex networks are far from being random combinations of the different layers, but their structures were found to be determined by hidden geometric correlations~\cite{kleineberg16}.

\subsection{Edge properties}

Due to the presence of multiple layers, a pair of nodes $(i,j)$ can be
connected through several edges. Given two layers $\alpha$ and $\beta$,
the edge overlap of the pair $(i,j)$~\cite{bianconi13,battiston14} is
defined as
\begin{equation}
o_{ij}^{[\alpha,\beta]} = \frac{a_{ij}^{[\alpha]} + a_{ij}^{[\beta]}}{2},
\end{equation}
where $o_{ij}^{[\alpha,\beta]}=1$ if i and j are connected at both
layers, $o_{ij}^{[\alpha,\beta]}=1/2$ if they are connected at one
layer only, and $o_{ij}^{[\alpha,\beta]}=0$ if the two nodes are node
connected.  For a generic number of layers $M$, the edge overlap is
defined as
\begin{equation}
o_{ij} =\frac{1}{M} \sum_{\alpha}a_{ij}^{[\alpha]}.
\end{equation}
This measure can be easily extended to the whole network as
\begin{equation}
o = \frac{2}{N(N-1)}\sum_{i, j \neq i} o_{ij},
\end{equation}
where the average is computed over all possible pairs of
nodes~\cite{battiston14}, or instead as
\begin{equation}
\omega =\frac{\sum_i\sum_{j>i} a_{ij}\lay{\alpha} }{M\sum_i \sum_j 1-
  \delta_{0, \sum_{\alpha} a_{ij}\lay{\alpha}}}
\end{equation}
where the average is restricted to the pairs of nodes which share at
least one edge~\cite{lacasa15}. Alternative definitions for the local
edge overlap on a node $i$ and the total overlap of two layers are
suggested in~\cite{bianconi13} and respectively read
\begin{equation}
\tilde o_{i}^{[\alpha,\beta]} = \sum_{j=1}^N \tilde
o_{ij}^{[\alpha,\beta]} = \sum_{j=1}^N a_{ij}^{[\alpha]}
a_{ij}^{[\beta]},
\end{equation}
and
\begin{equation}
\tilde o^{[\alpha,\beta]} = \sum_{i<j} a_{ij}^{[\alpha]} a_{ij}^{[\beta]},
\end{equation}
where $\tilde o_{ij}^{[\alpha,\beta]}=1$ when both layers have a link
between $i$ and $j$ and $\tilde o_{ij}^{[\alpha,\beta]}=0$ otherwise.
In the same spirit, a similar measure of edge correlations is the
so-called multiplexity~\cite{gemmetto15}, defined as
\begin{equation}
m\lay{\alpha,\beta}=\frac{2 \sum_{i<j} \rm
  {min}(a_{ij}\lay{\alpha}a_{ij}\lay{\beta})}{K\lay{\alpha}+K\lay{\beta}}
\end{equation} 
where $K\lay{\alpha}$ ($K\lay{\beta}$) is the total number of edges at
layer $\alpha$ ($\beta$). Notice that $m\lay{\alpha,\beta}$ takes
values in the range $[0,1]$.

A somehow dual quantity is the so-called edge intersection index,
\begin{equation}
INT = M \frac{\sum_{i,j=1}^{N}\min_{\alpha}\left\{a_{ij}\lay{1},
    a_{ij}\lay{2}, \ldots,
    a_{ij}\lay{M}\right\}}{\sum_{\alpha=1}^{M}\sum_{i,j=1}^{N}a_{ij}\lay{\alpha}}
\end{equation}
which measures the probability of finding a pair of nodes that is
connected by an edge on all the $M$ layers of the
multiplex~\cite{dedomenico15red}.

An alternative characterisation of edge correlations can be based on
the conditional probability to find a link at layer $\alpha$ given the
presence of an edge between the same nodes at layer
$\beta$~\cite{battiston14}
\begin{equation}
P(a_{ij}^{[\alpha]}|a_{ij}^{[\beta]})=\frac{\sum_{ij}a_{ij}^{[\alpha]}a_{ij}^{[\beta]}}{\sum_{ij}a_{ij}^{[\beta]}}
\label{eq:prob_cond}
\end{equation}
If layer $\beta$ has weighted edges, it is also possible to look at
the conditional probability ${P^{\rm
    w}}(a_{ij}^{[\alpha]}|w_{ij}^{[\beta]})$ to have a link at layer
$\alpha$ given its weight on layer $\beta$.  If ${P^{\rm w}}$ shows an
increasing trend as a function of $w$, this phenomenon goes under the
name of edge reinforcement, since a stronger link on one layer implies
a higher chance to find the same edge on a different
layer~\cite{battiston14}.

\subsection{Mesoscale properties}

Complex networks are usually characterised by non-trivial structural
patterns not only at the level of single-node properties but also, and
more importantly, at the level of sub-graphs. A lot of attention has
been devoted to the analysis of statistically significant sub-graphs in
single-layer networks, also known as motifs. It has been found that a
few specific sub-graphs are over-represented in real systems compared to
their abundance in equivalent networks obtained by randomising the
original graph~\cite{Alon02,Alon04}. Due to the additional level of
richness provided by the layered structure of multiplex networks, the
multilink, i.e. the organisation of the edges between the same pair of
nodes $(i,j)$ across the $M$ layers, is the most basic
motif~\cite{bianconi13,menichetti14}. Similarly, m-triads and
m-triangles used for the definition of node clustering coefficients
are multiplex motifs~\cite{battiston14}. The problem of isomorphisms in multi-layer is studied in Ref.~\cite{kivela15}. A general classification in three levels of higher-order
motifs is presented in Ref.~\cite{battiston16brain}: at the first level connected subgraphs are distinguished according to their number of nodes; at the second level different patterns are classified on the aggregated network and eventually, at the third level, the exact multiplex connectivity pattern is identified. Such approach has been used for the analysis of two-layer networks based on structural and
functional connectivity in the human brain~\cite{battiston16brain}.

Another remarkable feature of networked systems is the tendency of
their units to cluster together in tightly-knit groups, giving rise
to non-trivial community structures. Communities are also observed in
multiplex networks, even if there is not to date an agreed definition
of what a multiplex community is~\cite{mucha2010}. Some of the efforts
in the characterisation of the communities of a multiplex have been
focused on the quantification of the similarities in the community
structure observed at different layers. In general, given two layers
$\alpha$ and $\beta$ and their partitions in communities ${\cal
  P}_{\cal \alpha}$ and ${\cal P}_{\cal \beta}$ , their similarity can
be measured through the normalised mutual information
(NMI)~\cite{dedomenico15comm,battiston16comm}
\begin{equation}
NMI({\cal P}_{\cal \alpha}, {\cal P}_{\cal \beta})  = \frac{-2\sum^{M_\alpha}_{m=1}\sum^{M_\beta}_{m'=1}
N_{m m'}\log\left(\frac{N_{m m'}N}{N_{m}N_{m'}}\right)}
{\sum^{M_\alpha}_{m=1}N_{m}\log\left(\frac{N_{m}}{N}\right)
 + \sum^{M_\beta}_{m'=1}N_{ m'} \log\left(\frac{N_{ m'}}{N}\right)}
\end{equation}
where $N_{m m'}$ is the number of nodes in common between community
$m$ of partition ${\cal P}_{\cal \alpha}$ and community $m'$ of
partition ${\cal P}_{\cal \beta}$, while $N_{m}$ and $N_{ m'}$ are
respectively the number nodes in the two communities $m$ and $m'$.  We
note that such measure was originally suggested to compare the
community structure obtained on the same single-layer networks from
different algorithms. A different similarity measure is suggested in
Ref.~\cite{iacovacci15}, in terms of the possibility to infer the
community structure at layer $\alpha$ using information about the
community structure at layer $\beta$.

The information about the decomposition in communities of different
layers can be combined together to define a multilayer partition in
communities. A classical approach is that described in
Ref.~\cite{mucha2010}, which proposed a generalisation of the concept of
modularity to multiplex, multi-slice, and temporal networks. Among the
genuinely multiplex methods to extract the community decomposition of
a system we find particularly interesting the approach proposed in
Ref.~\cite{dedomenico15comm}, which extends the Infomap
algorithm~\cite{rosvall08}, based on the minimisation of the
description length of a partition in communities, to the case of
multiplex networks.

%%%%%%%%%%%%%%%%                     MODELS             %%%%%%%%%%%%%%%%%%%%%
\section{Models of multiplex networks}
\label{sec:models}

The characterisation of the structure of a network is normally
accompanied by a modelling effort aiming at quantifying how special or
peculiar are the observed patterns, e.g. in terms of how probable is
to find them in an appropriately chosen family of random graphs, and
which are the mechanisms that determine their appearance. In the
following we review two classes of models of multiplex networks,
namely static random graph models and growing networks.

%%%%%%%% Ensembles
\subsection{Microcanonical and Canonical Ensembles}

A standard approach to study the structure of a given network is to
quantify how probable is to observe a network with similar properties
in an appropriately defined ensemble of random graphs whose elements
satisfy certain constraints. For instance, it is a well-known fact
that graphs with power-law degree distributions are extremely rare in
the classical Erd\"os-Renyi random graph ensemble, where each pair of
nodes are connected with a given probability $p$. As a consequence,
the hypothesis that power-law degree distributions arise as a result
of a uniform distribution of edges across the nodes can be safely
rejected, and we can conclude that some other mechanism should be at
work in the formation of graphs with heterogeneous degree
sequences.

An ensemble of graphs is determined by a set of constraints that its
elements should satisfy. According to the type of constraints, we can
identify at least two classes of random network ensembles, namely
canonical ensembles, where each graph of the ensemble satisfies the
set of constrains on average (soft constraints), and microcanonical
ensembles, where each graph satisfies all the constraints exactly
(hard constraints). It is possible to define a sequence of canonical
and microcanonical ensembles of multiplex networks~\cite{bianconi13},
where the constraints are just the average degree at each of the $M$
layers, or the degree distribution of each layer, or the degree
distribution together with the distribution of edge overlap, and so
on.

Each multiplex networks ensemble is defined by providing the
probability $P(\bm A)$ for each of the possible configuration of
multiplex networks $\bm A$ which satisfy the constraints. Starting
from $P(\bm A)$, the Shannon entropy of the ensemble is defined as
\begin{equation}
S= - \sum_{\mathcal M} P(\mathcal M) \ln P(\mathcal M)
\end{equation}
~\cite{bianconi13}.

For the special case of uncorrelated multiplex networks, we have
\begin{equation}
\langle a_{ij}\lay{\alpha} a_{ij}\lay{\beta}  \rangle = \langle a_{ij}\lay{\alpha} \rangle \langle a_{ij}\lay{\beta}  \rangle
\end{equation}
the probability $P(\mathcal M) \equiv P(\bm A)$ can be factorised into
the probability of observing each single layer, i.e.
\begin{equation}
P(\bm A )= \prod_{\alpha=1}^M P\lay{\alpha} (A\lay{\alpha}).
\end{equation}
In this particular case, the entropy of the multiplex ensemble reads
\begin{equation}
S = \sum_{\alpha=1}^M S\lay \alpha = - \sum_{\alpha=1}^M P\lay{\alpha} (A\lay{\alpha}) \ln (P\lay{\alpha} (A\lay{\alpha})).
\end{equation}

In the following we focus on the canonical - indicated by $C$ - and
microcanonical - denoted by $M$ - ensembles of multiplex networks.
Let us assume that we have $T$ soft constraints such that
\begin{equation}
\label{eq:constraints}
\sum_{\bm A} P(\mathcal M) F_{\mu}(\mathcal M) = C_{\mu}
\end{equation}
where $\mu = 1, \ldots, T$, and $F_{\mu}(\mathcal M)$ describes how
such constraints are imposed on the network, such as the degree of
each node of the network at each layer $\alpha$, or the total number
of edges $K\lay{\alpha}$ for $\alpha=1, \ldots, M$. The probability
$P_C(\mathcal M)$ of observing the multiplex $\mathcal{M}$ can be
obtained by maximising the entropy $S$ under the given set of
constraints. By solving the optimisation problem one obtains:
\begin{equation}
P_C(\mathcal M)= \frac{1}{Z_C} \rm{exp} \biggl [ - \sum_{\mu} \lambda_{\mu} F_{\mu} (\mathcal M)    \biggr ]
\end{equation}
where $Z_C$ is the partition function of the canonical multiplex
ensemble and the values of the Lagrangian multipliers $\lambda_{\mu}$
are obtained by satisfying Eq.\ref{eq:constraints} imposing such
functional form for $P_C(\mathcal M)$. In the canonical multiplex
ensembles we have:
\begin{equation}
S= \sum_{\mu} \lambda_{\mu} C_{\mu} + \ln Z_c.
\end{equation}

Conversely, in the microcanonical multiplex ensemble each multiplex
configuration compatible with the hard constraints has the same
probability
\begin{equation}
P_M(\mathcal M) = \frac{1}{Z_M} \prod^{T}_{\mu=1} \delta [F_{\mu}(\mathcal M),C_{\mu}] 
\end{equation}
where $\delta$ is the Kronecker delta function and $Z_M=\sum_{\bm
  A}\prod^{T}_{\mu=1} \delta [F_{\mu}(\mathcal M),C_{\mu}] $ is the
microcanonical partition function, accounting for the number of
multiplex networks satisfying the $T$ hard constraints
$F_{\mu}(\mathcal M)=C_{\mu}$.  By defining the entropy of these
ensembles as $N \Sigma$, such entropy reads
\begin{equation}
N \Sigma = \sum_{\bm A} P_M(\mathcal M) \ln P_M(\mathcal M) = \ln Z_M
\end{equation}
where $\Sigma$ is the Gibbs entropy of the multiplex ensemble. It can
be shown that the Gibbs entropy $\Sigma$ is related to the
corresponding Shannon entropy $S$ by $N \Sigma = S - N \Omega$, where
$\Omega$ is the logarithm of the probability that in the related
canonical multiplex ensemble the hard constraints $F_{\mu}(\mathcal
M)$ are satisfied.

The author of Ref.~\cite{bianconi13} provided an exhaustive
explanation of how the entropy and the partition function can be
computed in different classes of multiplex networks with increasingly
stringent sets of constraints, both for the canonical and for the
microcanonical ensembles. The same approach has been generalised to a
number of more complicated structures, including spatial multiplex
networks~\cite{halu14} and multiplex networks with heterogeneous
activities of the nodes~\cite{cellai16}. The Authors of Ref.~\cite{sagarra15} study the canonical ensemble of the overlapping networks generated by merging different layers, where information on the connection between nodes is only accessible at the aggregated level.

%%%%%%%% Null models
\subsection{Models of node and layer activity}

The concept of node and layer activity is peculiar to multilayer
networks, and it is interesting to assess whether simple models can
give account for the observed heterogeneous distributions of node and
layer activities. In the following we provide a brief review of some
null models proposed so far to quantify the peculiarity of given
distributions of node and layer activities.

Let us consider two layers $\alpha$ and $\beta$ with $N\lay{\alpha}$
and $N\lay{\beta}$ active nodes respectively. If initially the two
layers have no active nodes and we then sample uniformly at random
from $\{1, 2, 3, \ldots, N\}$ $N\lay{\alpha}$ nodes on layer $\alpha$
and $N\lay{\beta}$ nodes on layer $\beta$ and we activate them, then
the probability that $m$ of them are active at both layers follows a
hypergeometric distribution
\begin{equation}
p(m; N, N\lay{\alpha}, N\lay{\beta})=\frac{  \binom{N\lay{\alpha}}{m}  \binom{N-N\lay{\alpha}}{N\lay{\beta} - m}    }{\binom{N}{N\lay{\beta}}},
\end{equation}
according to which the expected number of nodes active at both layers
is equal to $N\lay{\alpha}N\lay{\beta}/N$, the expected pairwise
multiplexity is
\begin{equation}
  \tilde Q\lay{\alpha, \beta}=\frac{N\lay{\alpha}N\lay{\beta}}{N^2}
  \label{eq:multiplexity}
\end{equation}
and the expected Hamming distance reads
\begin{equation}
\tilde H\lay{\alpha, \beta}= \frac{\sum_{m=0}^{N\lay{\beta}}
  (N\lay{\alpha} + N\lay{\beta} - 2m) \times p(m; N, N\lay{\alpha},
  N\lay{\beta}) }{\rm{min}(N, N\lay{\alpha}+N\lay{\beta})}
\label{eq:hamming}
\end{equation}
This is the simplest model of node activation and is known as the
hypergeometric model~\cite{nicosia15}. However, the authors of
Ref.~\cite{nicosia15} have shown that the distribution of pairwise
multiplexity and pairwise Hamming distance in real-world multiplex
networks is not compatible with those given in
Eq.~\ref{eq:multiplexity} and Eq.~\ref{eq:hamming}.

Let us now consider the problem of constructing a multiplex networks
with a fixed number of layers $M$, a fixed number of nodes $N$ which
are active on at least one of the $M$ layers, and where each node $i$
has an assigned node activity $B_i$, which is for instance set equal
to the node activity observed in the real network. By sampling for
each node one of the $\binom{M}{B_i}$ vectors of node-activity with
$B_i$ non-zero entries, the distribution of the total node activity of
the original system is kept fixed, whereas the correlations in the
layer activity and the distribution of the node-activity vectors are
not preserved. Moreover, in such a model all layers have the same
expected number of active nodes:
\begin{equation}
\tilde N\lay{\alpha}=\frac{1}{M}\sum_i B_i.
\end{equation}
This model is known as the multi-activity deterministic
model~\cite{nicosia15}.  A variation of the model is constructed by
activating node $i$ in each layer $\alpha$ with probability $\bar
B_i=B_i/M$, so that the expected activity of each layer stays the same
but the original node-activity distribution is not preserved. This
model is known as the multi-activity stochastic
model~\cite{nicosia15}.

Finally, it is possible to construct a model for a 2-layer multiplex
network where the degree distributions of each layer is kept fixed,
and where one can control the edge overlap $\omega$ by rewiring a
certain fraction $r$ of the edges. The model was introduced in
Ref.~\cite{diakonova16}. For simplicity, let us assume that the two
layers have the same number of edges
$K\lay{\alpha}=K\lay{\beta}=K$. If we start from two identical
networks, we have maximum edge overlap $\omega=1$. If we now keep
fixed the structure on one of the two layers, and rewire one of the
edges of the other layer, the number of links present in both layers
decreases by one unit, while the number of those present in only one
of the two layers increases by two units. Consequently, if we rewire a
fraction $r$ of the $K$ edges of the second layer in such a way that
each rewire decreases the number of edges existing on both layers, we
obtain:
\begin{equation}
\omega = \frac{(1-r)K}{(1+r)K} = \frac{(1-r)}{(1+r)} 
\end{equation}
By inverting such relation, we find that a given overlap $\omega$
corresponds to a rewire $r$ equal to $r=(1-\omega)/(1+\omega)$. In
practice, this model allows to obtain a prescribed value of edge
overlap by rewiring a certain fraction $r$ of the edges in one of the
two layers.

%%%%%%%% Growth models
\subsection{Growth models of multiplex networks}

In this section we review a few growth models for multiplex networks.
The most simple example of this class is a model of layer-growth,
aimed at explaining the fat-tail distribution of layer activity
observed in empirical data~\cite{nicosia15}. The model works as
follows.  We start at time $t_{0}=0$ with a multiplex with $M_{0}$
layers and $N$ nodes. At each time $t$, a new layer $\alpha$ joins the
network with $N\lay{\alpha}$ nodes to be activated, where
$N\lay{\alpha}$ can be observed from the data-set we are attempting to
reproduce. Each node $i$ has then a probability to be active on that
layer at time $t$ equal to:
\begin{equation}
p_i(t)=A+B_i(t),
\end{equation}
where $B_i(t)$ is the number of layers where node $i$ is already
active and $A>0$ is a constant that allows the activation of nodes not
yet active in the multiplex. When the number of layers in the model
increases, the distribution of layer activity $P(N\lay{\alpha})$
approaches a power law.

Another important class of growth models is that where not layers, but
individual nodes join sequentially the network, for instance by
connecting to preexisting vertices on possibly different layers. In
such regard, it is clear that the specific shape of the attachment
function determines the long-term statistical properties of the final
multiplex graph. In single-layer networks, a particularly well-studied
case is the so-called preferential attachment, where nodes choose to
attach to older vertices depending on a function (in the simplest case
linear) of their degree $k$. In a multiplex network the degree of each
node $j$ is a vector and the probability $\Pi_{i \to j}\lay{\alpha}$
that a new node $i$ attaches to $j$ on a given layer $\alpha$ in
general depends on all its components. In formula:
\begin{equation}
  \rm {\Pi}_{i \to j}\lay{\alpha}=\frac{F_j\lay{\alpha}({\bold
      k}_j)}{\sum_\ell F_\ell \lay{\alpha}({\bold k}_\ell)}
  \label{eq:attach}
\end{equation}

The most simple class of preferential attachment models is obtained by
considering linear attachment kernels, i.e. by setting
$F\lay{\alpha}_j$ as a convex combination of the degrees of node $j$
at all the layers~\cite{nicosia13,kim13}. The interesting result is
that linear attachment kernels produce multiplex networks whose layers
have power-law degree distributions, but where inter-layer degree
correlations are always positive, meaning that a hub on one layer is
also a hub on the other layer as well. This is due to the fact that
the expected final degree of a node on a certain layer is determined
solely by the time at which it joins the network~\cite{nicosia13}. A
generalisation of the closed-form solutions for the joint degree
distribution of heterogeneously growing multiplex networks with
arbitrary number of layers and arbitrary times can be found
in~\cite{momeni15}.

A more interesting class of multiplex networks is obtained by
considering non-linear attachment kernels. The authors of
Ref.~\cite{nicosia14} started from the case of multiplex networks with
two layers, using the attachment kernel:
\begin{equation}
  F\lay{1}_j \propto \left(k\lay{1}_j\right)^{\alpha}
  \left(k\lay{2}_j\right)^{\beta}
\end{equation}
where $\alpha, \beta\in \mathbb{R}$. By tuning the relative values of
the exponents $\alpha$ and $\beta$, one can obtain multiplex networks
where each layer has either an exponential, a power-law, or a
condensed degree distribution (where super-hubs with extensive degrees
appear). Moreover, with non-linearity it is possible to get non only positive inter-layer degree correlations as in the case of the linear model, but also null and negative correlations that have been observed in real-world systems~\cite{nicosia15}. In the same work the authors suggested
several possible generalisations of the model to the case of multiplex
networks with $M\le 2$ layers. An interesting model of multiplex
network growth which takes into account weighted links, aimed at
reproducing the structure of some layered social networks, can be
found in Ref.~\cite{murase14}.

\subsection{Models of multiplex communities}

Simple preferential attachment models, while able to reproduce some
empirical patterns such as inter-layer degree correlations, do not
allow to construct multiplex networks with strong community
structure. More sophisticated models able to produce tunable
intra-layer and inter-layer community structure have been suggested,
based on intra-layer and inter-layer triadic closure mechanisms on
2-layer multiplexes~\cite{battiston16comm}. In that model a node $i$
arrives and selects one of the layers at random, and a node $n_1$ in
that layer as its first neighbour. The following $m-1$ links on the
same layer will be to a neighbour of $n_1$ with probability $p$, or to
a uniformly sampled node in the same layer with probability
$1-p$. Once $m$ links have been created on the first layer, node $i$
starts creating links on the other layer. In particular, the first
edge on the other layer is created with probability $p^*$ to the same
node $n_1$, and at to a node sampled uniformly at random with
probability $1-p^*$. The remaining links on the second layer are
placed again with probability $p$ and $1-p$. In this model, the value
of the parameter $p$ determines the strength of communities on each
layer, with higher values of $p$ corresponding to tighter communities,
while $p^*$ tunes the extent of overlap (i.e., number of shared nodes)
between communities in different layers. Interestingly, the model was
able to reproduce some of salient characteristics of multiplex
collaboration networks.

We note here that multiplex networks with given community structure can also be generated through stochastic block models~\cite{peixoto15}, and that in some cases aggregated networks can be better fitted by multilayer block models, hinting at the existence of different levels in the considered data~\cite{valles16}.

\section{Conclusions}

Networks are responsible for the emergence of a variety of complex behaviours in social, economical, technological and biological systems, and multilayer networks are the last frontier of research in this field. The theory of multiplex networks has already proven quite successful in modelling the structure of intrinsically multidimensional relational systems, showing at the same time that the presence of more than a single type of interaction is responsible for new levels of complexity. The advances made in this field in the last few years are definitely encouraging, and there is still a lot of open problems to address in depth and many questions still waiting to be asked. We strongly believe that multiplex networks are an extremely active and interesting area of research, and we really hope that this brief review will contribute to spur the curiosity of researchers who are interested in studying the structure of real-world systems.


\begin{thebibliography}{99}

%1
\bibitem{barabasi02} R. Albert, A.L. Barab\'asi, Reviews of Modern
  Physics \textbf{74}, 47 (2002)

\bibitem{newman03} M. E. J. Newman, SIAM Review \textbf{45}, 167-256 (2003)
  
\bibitem{Boccaletti2006} S. Boccaletti, V. Latora, Y. Moreno,
  M. Chavez, D. U. Hwang, Physics Reports \textbf{424}, 175-308 (2006)
  
\bibitem{newman10} M. E. J. Newman, Networks: an Introduction (Oxford
  University Press, Oxford, UK) (2010)

%5
\bibitem{strogatz01} S. Strogatz, Nature \textbf{410}, 268-276 (2001)

\bibitem{arenas08} A. Arenas et al., Physics Reports
  \textbf{469} (3), 93-153 (2008)

\bibitem{bullmore09} E. Bullmore, O. Sporns, Nature Reviews on
  Neuroscience \textbf{10}, 186-198 (2009)
  
\bibitem{castellano09} C. Castellano, S. Fortunato, V. Loreto, Reviews
  of Modern Physics \textbf{81}, 591-646 (2009)

\bibitem{barthelemy11} M. Barth\'elemy, Physics Reports \textbf{499}, 1-101
  (2011)

%10  
\bibitem{fallani14} F. De Vico Fallani, J. Richiardi, M. Chavez,
  S. Achard, Philosophical Transactions of the Royal Society of London
  B: Biological Sciences, \textbf{369} (2014)
 
\bibitem{boccaletti14} S. Boccaletti et al., Physics Reports
  \textbf{544} 1 1-122 (2014) 

\bibitem{kivela14} M Kivela et al., Journal of Complex Networks \textbf{2} (3)
   203 - 271 (2014)

\bibitem{lee15} K.-M. Lee, B. Mina, K.-I. Goh, Eur. Phys. J. B \textbf{88}, 48
  (2015)
 
\bibitem{battiston14} F. Battiston, V. Nicosia, V. Latora, Physical
  Review E \textbf{89} 032804 (2014)

%15
\bibitem{menichetti14} G. Menichetti et al., PLoS ONE \textbf{9} (6):
  e97857. doi:10.1371/journal.pone.0097857 (2014) 
  
\bibitem{dedomenico13} M. De Domenico et al., Physical Review X \textbf{3} (4),
  041022 (2013)

\bibitem{superdiffusion} S. G\'omez et al., Physical Review Letters
  \textbf{110}, 028701 (2013)
  
\bibitem{radicchi14} F. Radicchi, A. Arenas, Nature Physics, \textbf{9}, 717
  (2013)
  
\bibitem{nicosia15} V. Nicosia, V. Latora, Physical Review E
  \textbf{92} 032805 (2015)

%20
\bibitem{cellai16} D. Cellai, G. Bianconi, Physical Review E \textbf{93},
  032302 (2016)
  
\bibitem{cozzo15} E. Cozzo et al., New Journal of Physics \textbf{17} (7),
  073029 (2015)

\bibitem{sola13} L. Sol\'a et al., Chaos: An Interdisciplinary Journal
  of Nonlinear Science \textbf{23} (3), 033131 (2013)
  
\bibitem{battiston16rw} F. Battiston, V. Nicosia, V. Latora, New
  Journal of Physics \textbf{18} (4), 043035 (2016)

\bibitem{sole16rw} A. Sol\'e-Ribalta, M. De Domenico, S. G\'omez and
  A. Arenas, Physica D \textbf{323}, 73-79 (2016)

\bibitem{halu13} A. Halu, R. J. Mondrag\'on, P. Panzarasa,
  G. Bianconi, PLoS ONE, \textbf{8} (10):
  e78293. doi:10.1371/journal.pone.0078293 (2013)

\bibitem{morris12} M. G. Morris, M. Barthelemy, Physical Review
  Letters \textbf{109} 12 128703 (2012)

\bibitem{nicosia13} V. Nicosia, G. Bianconi, V. Latora, M. Barthelemy,
  Physical Review Letters \textbf{111} 058701 (2013)

\bibitem{aleta16} A. Aleta, S. Meloni, Y. Moreno, arXiv preprint, arXiv: 1607.00072 (2016)
  
\bibitem{dedomenico15vers} M. De Domenico, A. Sol\'e-Ribalta,
  E. Omodei, S. G\'omez, A. Arenas, Nature Communications \textbf{6}
  (2015)

%30
\bibitem{sole14} A. Sol\'e-Ribalta, M. De Domenico, S. G\'omez, A.
  Arenas, Proceedings of the 2014 ACM conference on Web Science
  149-155 (2014)

\bibitem{Pastor2001} R. Pastor-Satorras, A. Vazquez, A. Vespignani,
  Physical Review Letters \textbf{87}, 258701 (2001)
  
\bibitem{lacasa15} L. Lacasa, V. Nicosia, V. Latora, Scientific
  reports \textbf{5} (2015)

\bibitem{ferraz15} G. Ferraz de Arruda, E. Cozzo, Y. Moreno,
  F. A. Rodrigues, arXiv preprint, arXiv:1507.04550 (2015)

\bibitem{sole2013} A. Sol\'e-Ribalta et al., Physical Review E \textbf{88} (3),
  032807 (2013)

\bibitem{sanchez2014} R.J. S\'anchez-Garc\'ia, E. Cozzo, Y. Moreno, Physical
  Review E \textbf{89} (5), 052815 (2014)

\bibitem{dedomenico15red} M. De Domenico, V. Nicosia, A. Arenas,
  V. Latora, Nature Communications \textbf{6} (2015)

%35
\bibitem{kleineberg16} K.-K. Kleineberg, M. Boguna, M. Angeles Serrano, F. Papadopoulos, Nature Physics
  doi:10.1038/nphys3812 (2016)

\bibitem{bianconi13} G. Bianconi, Physical Review E \textbf{87} 6
  062806 (2013)

\bibitem{gemmetto15} V. Gemmetto, D. Garlaschelli, Scientific Reports
  \textbf{5} 9120 (2015)

\bibitem{Alon02} R. Milo et al., Science \textbf{298}, 824--827 (2002)

\bibitem{Alon04} R. Milo et al., Science \textbf{303}, 1538--1542 (2004)

%40    
\bibitem{kivela15} M. Kivela, M.A. Porter, arXiv preprint
  arXiv:1506.00508 (2015)

\bibitem{battiston16brain} F. Battiston, M. Chavez, V. Nicosia,
  V. Latora, arXiv preprint arXiv:1606.09115 (2016)

\bibitem{mucha2010} P. J. Mucha et al., Science \textbf{328}, 876--878 (2010)
  
\bibitem{dedomenico15comm} M De Domenico, A Lancichinetti, A Arenas, M
  Rosvall, Physical Review X \textbf{5} (1), 011027 (2015)

\bibitem{battiston16comm} F. Battiston, J. Iacovacci, V. Nicosia,
  G. Bianconi, V. Latora, PLOS ONE \textbf{11} (1) e0147451 (2016)

%45  
\bibitem{iacovacci15} J. Iacovacci, Z. Wu, G. Bianconi, Physical Review E
  \textbf{92} (4), 042806 (2015)

\bibitem{rosvall08} M. Rosvall, C. T. Bergstrom, Proceedings of the
  National Academy of Sciences of the U.S.A. \textbf{105}, 1118-1123 (2008)

\bibitem{halu14} A. Halu, S. Mukherjee, G. Bianconi, Physical Review E
  \textbf{89}, 012806 (2014)

\bibitem{sagarra15} O. Sagarra, C. J. Perez Vicente, A. Diaz-Guilera, Physical Review E
  \textbf{92} (5), 052816 (2015)

\bibitem{diakonova16} M. Diakonova, V. Nicosia, V. Latora, M. San
  Miguel, New J. Phys. \textbf{18} 023010 (2016)

\bibitem{kim13} J, Y. Kim and K.-I. Goh, Physical Review Letters \textbf{111}, 058702
  (2013).

\bibitem{momeni15} N. Momeni, B. Fotouhi, Physical Review E \textbf{92}, 062812 (2015)

\bibitem{nicosia14} V. Nicosia, G. Bianconi, V. Latora, M. Barthelemy,
  Physical Review E \textbf{90}, 042807 (2014)

\bibitem{murase14} Y. Murase, J. T\"or\"ok, H.-H. Jo, K. Kaski,
  J. Kert\'esz, Physical Review E \textbf{90}, 052810 (2014)

\bibitem{peixoto15} T. P. Peixoto, Physical Review E \textbf{92}, 042807 (2015)

\bibitem{valles16} T. Valles-Catala, F. A. Massucci, R. Guimera,  M. Sales-Pardo, Physical Review X \textbf{6}, 011036 (2016)



\end{thebibliography}
\end{document}